\documentclass[prl,amsmath,amssymb,twocolumn,superscriptaddress]{revtex4-1}

\usepackage{amsmath}    
\usepackage{amsfonts}
\usepackage{amssymb}
\usepackage{graphicx}   
\usepackage{dcolumn}
\usepackage{bm}

\begin{document}

\title{Collisionless Weibel shocks: full formation mechanism and timing}

\author{A. Bret}
\affiliation{ETSI Industriales, Universidad de Castilla-La Mancha, 13071 Ciudad Real, Spain}
 \affiliation{Instituto de Investigaciones Energ\'{e}ticas y Aplicaciones Industriales, Campus Universitario de Ciudad Real,  13071 Ciudad Real, Spain.}

 \author{A. Stockem}
\affiliation{GoLP/Instituto de Plasmas e Fus\~{a}o Nuclear, Instituto Superior T\'{e}cnico, Universidade de Lisboa, Lisbon, Portugal}
\affiliation{Institut f\"{u}r Theoretische Physik, Lehrstuhl IV: Weltraum- und Astrophysik, Ruhr-Universit\"{a}t Bochum, D-44780 Bochum, Germany}

 \author{R. Narayan}
\affiliation{Harvard-Smithsonian Center for Astrophysics,
60 Garden Street, MS-51 Cambridge, MA 02138, USA}

 \author{L.O. Silva}
\affiliation{GoLP/Instituto de Plasmas e Fus\~{a}o Nuclear, Instituto Superior T\'{e}cnico, Universidade de Lisboa, Lisbon, Portugal}

\date{\today }

\begin{abstract}
Collisionless shocks in plasmas play an important role in space
physics (Earth's bow shock) and astrophysics (supernova remnants,
relativistic jets, gamma-ray bursts, high energy cosmic rays). While
the formation of a fluid shock through the steepening of a large
amplitude sound wave has been understood for long, there is currently
no detailed picture of the mechanism responsible for the formation of
a collisionless shock. We unravel the physical
mechanism at work and show that an electromagnetic Weibel shock always forms when  two relativistic collisionless, initially unmagnetized, plasma shells encounter. The predicted shock formation time is in good agreement with 2D and 3D particle-in-cell simulations of counterstreaming pair plasmas. By predicting the shock formation time, experimental setups aiming at producing such shocks can be optimised to favourable conditions.
\end{abstract}


\maketitle

\section{Introduction}
In a fluid shock, when the upstream flow slows
down at the shock front, particles dissipate their bulk kinetic energy
through collisions among one another. As a result, the width of the
shock front is of the order of the collisional mean free path
\cite{Zeldovich}.

Earth's bow shock in the solar wind behaves very differently. Here,
the width of the shock front is about 100\,km \cite{PRLBow1,PRLBow2},
while the mean free path due to Coulomb collisions is a million times
larger, of order the Sun-Earth distance \cite{kasper2008}.  Earth's
bow shock and others like it are labelled ``collisionless'', since
they develop in media where the collisional mean free path is much
larger than the size of the system. Collisionless shocks are
ubiquitous in astrophysical environments and have the ability to
accelerate particles to high energies with power-law distributions
\cite{Blandford78,Bell1978a,Bell1978b,Spitkovsky2008a,Martins2009}.
Because of their likely role in generating high energy cosmic rays and
non-thermal radiation from sources like supernova remnants,
relativistic jets and gamma ray bursts \cite{Blandford1987,Piran2004},
collisionless shocks are currently the object of active research.

The theoretical foundations of collisionless
shocks in plasmas date back to the pioneering work of Sagdeev in the
1960's \cite{Sagdeev66}. More recently, kinetic computer simulations could capture their formation from the ab-initio collision of two plasma shells
\cite{Spitkovsky2005,kato2007,Chang2008,SilvaApJ}. However, while fluid shock
formation was understood long ago, e.g., through the steepening of a
large amplitude sound wave \cite{Zeldovich}, collisionless shock
formation still lacks a first principle understanding. The first stage of the formation, namely, the seeding of  electromagnetic instabilities, has been investigated in detail \cite{Chang2008,BretPoP2013,Medvedev1999}. But the subsequent stage of energy dissipation is still not yet understood, despite the tremendous insights provided by numerical simulations. Filling this gap would be relevant not only from the purely theoretical point of view, but it could be beneficial for optimising the experimental conditions for producing these shocks in the laboratory \cite{ross2012,Kugland2012,Fox2013,Huntington2013,Muggli2013,Sarri2013}.

Typically, in these experiments, two counterstreaming beams are generated by the irradiation of two opposing solid targets with a high-intensity laser pulse. This will give rise to ion-driven Weibel instability; the self-generated electromagnetic fields are probed, e.g., with ultrafast proton radiography techniques. The challenge with driving these shocks in the laboratory is achieving a forward directed beam rather than an isotropic heating of the target particles. The latter can give rise to so-called electrostatic shocks, showing a notable potential jump at the shock front, while Weibel-mediated shocks do not exhibit such jump \cite{Stockem2013}. An accurate understanding of the generation of both kinds of shocks is currently under scrutiny, and it seems that electrostatic shocks develop rather in non relativistic settings, whereas Weibel shocks involve relativistic velocities \cite{Stockem2013}. Our interest here is on the
relativistic Weibel shock.

In this article, the shock formation process is investigated in an idealized approach, for the collision of two initially unmagnetized symmetric cold pair plasmas. This reduces the number of free parameters, with the upstream Lorentz factor as the only free parameter. As the shells overlap, the counter-streaming plasma system formed in the overlapping region turns unstable. An analysis of the unstable spectrum for the system considered here shows the current-filamentation (Weibel-type) instability \cite{Weibel,Fried1959,DeutschPRE2005} is the fastest growing one \cite{BretPoPReview}. The instability picks up on the spontaneous magnetic fluctuations and amplifies them up to the level where it saturates. Knowing the growth rate, together with the level of fluctuations and the field at saturation, it is possible to derive an expression for the saturation time, $\tau_s \propto (\ln\gamma_0) /\delta $ [Eq.~(\ref{eq:sat-time}) below] with upstream Lorentz factor $\gamma_0$ and growth rate $\delta$ \cite{BretPoP2013}.

We are here concerned with the processes happening after
saturation. This question is of interest because, as we show, a shock
has not yet formed at $t=\tau_s$. The density jump between the
upstream plasma and the unstable plasma in the overlapping region is
still only a factor of 2, far short of the jump we expect in a shock.
For $t>\tau_s$, other non-linear processes must step in and
cause the system to form a proper shock.  Our goal here is to estimate
the shock formation time $\tau_f$ and to compare the theoretical
prediction with numerical simulations.

\begin{figure}[tb]
\begin{center}
\includegraphics[width=0.4\textwidth]{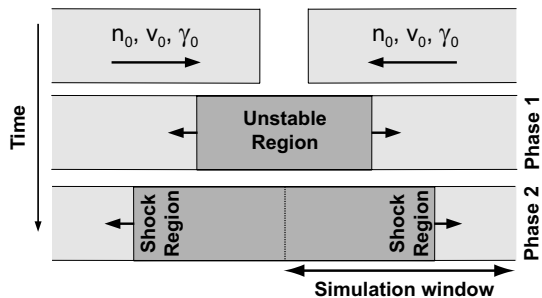}
\end{center}
\caption{Phases in the formation of a collisionless shock. Top: Two
  identical pair plasma shells approach each other. Middle (Phase 1):
  After the shells interpenetrate, the overlapping region turns
  unstable and the instability saturates. Bottom
  (Phase 2): Two shocks form near the border of each
  shell.} \label{fig:setup}
\end{figure}

\section{Saturation time and conditions at saturation}
In order to study the details of the full shock formation mechanism, we consider the  system pictured in Fig. \ref{fig:setup}.  In the relativistic regime ($v_0\sim c$), the
Weibel instability dominates in the overlapping region
\cite{BretPRL2008,BretPoPReview} and has a growth rate
\cite{Benford1973,califano3}
\begin{equation}\label{eq:gr}
  {\delta}={\omega_p}\, \sqrt{\frac{2}{\gamma_0}},
\end{equation}
where the plasma frequency $\omega_p$ is given by $\omega_p^2=4\pi n_0
q^2/m$, with $q$ and $m$ being the electron/positron charge and mass
respectively. For $1\ll\gamma_0\ll\mu \equiv mc^2/k_BT$, the
instability becomes non-linear and saturates at a time given by
\cite{BretPoP2013}
\begin{equation}\label{eq:sat-time}
\tau_s =\frac{1}{2 \delta} \,
\ln\left[
\frac{4}{15}\sqrt{\frac{6}{\pi}}n_0\left(\frac{c}{\omega_p}\right)^3 \sqrt{\gamma_0}  \mu \right]  \equiv  \frac{1}{2 \delta}\,\ln \Pi.
\end{equation}
At saturation, the perturbed magnetic field has an rms amplitude $B_s$
and the perturbations have a wave vector
$\mathbf{k}_b\perp\mathbf{v}_0$ with \cite{davidsonPIC1972,BretPoP2013}
\begin{equation}\label{eq:fieldsat}
B_s=\sqrt{\gamma_0}\,\frac{mc\omega_p}{q},~~~k_b=\frac{\omega_p}{c\sqrt{\gamma_0}}.
\end{equation}

In general, the fluctuations seeding the instability depend on the
shell history. In deriving the logarithmic factor in
Eq.~(\ref{eq:sat-time}), we used the spontaneous thermal fluctuations
which are present in a plasma \cite{Sitenko,Ruyer2013} even in the
absence of external perturbations. The result is therefore an upper
bound on the saturation time.

At the saturation time $t=\tau_s$, the instability has barely exited
the linear phase, so density perturbations in the overlapping shells
are still small. As a consequence, the density of the overlapping
region is just twice that of the individual shells.  This is confirmed
in particle-in-cell (PIC) simulations which we now
describe.

\begin{figure}[tb]
\begin{center}
\includegraphics[width=0.45\textwidth]{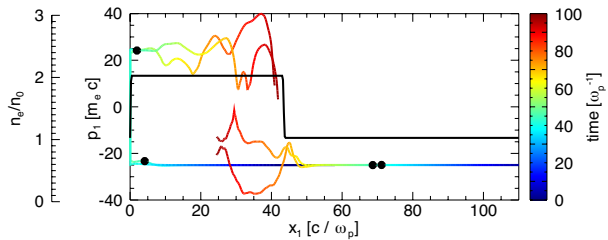}
\end{center}
\caption{(Color Online) Results from 2D PIC simulations for colliding
  cold plasma shells with $\gamma_0=25$. Compared to
  Fig.~\ref{fig:setup}, the simulation includes only the right half of
  the system, the left half being replaced by a reflecting wall at
  $x_1=0$. The $x_2$-integrated density at the saturation time
  $t=\tau_s=44\omega_p^{-1}$ is shown by the thick black curve, with
  the vertical scale indicated by the axis $n_e/n_0$ on the far
  left. Note that the density jump at the front of the overlapping
  region is only a factor of 2. The colored curves show the positions
  $x_1$ and momenta $p_1$ as a function of time for four particles,
  with the vertical scale indicated by the axis $p_1\,[m_ec]$.  The
  color code along the trajectories indicates time from $t=0$ (deep
  blue) up to $100\omega_p^{-1}$ (deep red), with $t=\tau_s$
  corresponding to the transition from light-blue to light-green
  (black dots).  Note that for $t<\tau_s$, all the trajectories are
  straight lines with a nearly constant $p_1 \sim -25 m_ec$
  (corresponding to $\gamma_0=25$).  For $t>\tau_s$, particles are
  decelerated and trapped by the magnetic field generated in the
  central region by the Weibel instability. As a result, particle
  trajectories in $x_1-p_1$ space begin to wander (note especially the
  orange and red segments).  The trapped plasma causes the density in
  the overlapping region to grow, leading to the formation of a shock
  at time $\tau_f \sim 2\tau_s =
  88\omega_p^{-1}$.} \label{fig:trap}
\end{figure}

We simulate only half the
system, replacing the other half by means of a reflecting wall at the
mid-point of the full symmetric system (Fig.~\ref{fig:trap}).  In the
simulations, the pair flow propagates towards the wall with a
relativistic Lorentz factor $\gamma_0$. The case shown in
Fig.~\ref{fig:trap} corresponds to $\gamma_0=25$, but we have run
other cases covering the range from $\gamma_0=10$ to $10^4$, as shown
later. The plasma is initially cold with \(\mu = 10^6\gamma_0\). The
two-dimensional simulation box spans a distance of $125 \,
\sqrt{\gamma_0} c/\omega_p$ along the propagation direction ($x_1$)
and $5 \, \sqrt{\gamma_0} c/\omega_p$ perpendicular to the flow
($x_2$), with a spatial resolution of \( \Delta x_1 = \Delta x_2 =
0.05 \, \sqrt{\gamma_0} c/\omega_p\).  The number of particles per
cell is 18 and the temporal resolution is \(\Delta t = \Delta x_1/2\).

For the particular simulation shown in Fig.~\ref{fig:trap}
($\gamma_0=25$), the saturation time $\tau_s=44\omega_p^{-1}$. The
solid black line shows the numerical density profile at this time. We
see that the density jump at the shock is almost exactly a factor of
2, and that there are practically no fluctuations in the integrated
density. This is consistent with the fact that at $t=\tau_s$ the
fluctuations due to the Weibel instability are just going non-linear.
In contrast, the density jump we expect for a fully formed 2D shock is a
factor of 3 [Eq. (\ref{eq:jump})].  Clearly, there must be a period of
time beyond $t=\tau_s$ when non-linear processes cause the system to
evolve further until, at some time $\tau_f>\tau_s$, a proper shock
with the correct jump conditions is formed (Phase 2, bottom panel, in
Fig.~\ref{fig:setup}). We are interested in estimating $\tau_f$.

\section{Penetration length and shock formation}
Up to time
$t=\tau_s$, incoming particles travel in nearly straight lines and are
hardly deflected at all, because the perturbations are still in the
linear stage and therefore small. This is confirmed in
Fig.~\ref{fig:trap} which shows trajectories of four selected
particles in the $x_1-p_1$ plane.  Right until
$t=\tau_s=44\omega_p^{-1}$, the particle momenta $p_1$ remain
practically constant (except for a reversal of the sign of $p_1$ for
two particles which are reflected at the wall).  One consequence of
the nearly free-streaming flow of the two shells is that, at
$t=\tau_s$, the half-length of the overlapping region is simply
(setting $v_0=c$)
\begin{equation}
L=c\tau_s = \frac{\ln \Pi}{2\sqrt{2}}\,\frac{1}{k_b}.
\end{equation}

At $t=\tau_s$, the overlapping region is filled with a fluctuating
magnetic field with strength and coherence scale given by
Eqs.~(\ref{eq:fieldsat}). This field is now strong enough to cause
significant deflections in the trajectories of particles entering the
overlapping zone.  The Larmor radius of an incoming particle of
Lorentz factor $\gamma_0$ in a uniform magnetic field $B_s$ is
\begin{equation}
r_L = \frac{\gamma_0 m c^2}{q B_s} = \frac{c\sqrt{\gamma_0}}{\omega_p}
=\frac{1}{k_b}.
\end{equation}
Since $r_L$ is of the order of the coherence length $1/k_b$ of the
fluctuating field, the mean free path of the particle, or the
isotropization length, is also of the same order. Furthermore, this
length is shorter than the half-length $L$ of the overlapping region
as long as $\ln\Pi > 2\sqrt{2}$, a condition that is always
satisfied for cold colliding plasmas.

The net consequence of the above statements is that, for $t>\tau_s$,
the incoming plasma shells no longer travel freely through each other
but are stopped and trapped in the overlapping region. This causes the
density to increase, resulting in a bona fide shock.

\section{Shock formation time}
The adiabatic index of a 2D
relativistic fluid is $\Gamma = 3/2$. Correspondingly, the expected
density ratio across a 2D relativistic shock is
\cite{blandford76,Stockem2012},
\begin{equation}\label{eq:jump}
\frac{n_d}{n_0}=1+\frac{\gamma_0+1}{\gamma_0 (\Gamma-1)} \sim 3
~~{\rm(in~2D)},
\end{equation}
where $n_d$ is the downstream density. Thus, between the saturation
time $\tau_s$ and the shock formation time $\tau_f$, the density in
the central region needs to increase from $2n_0$ to $3n_0$.  Now, the
size of the overlapping region will not increase much during this
period because incoming plasma is completely stopped and the
back-reaction due to the pressure in the overlapping region is
insufficient to balance the ram pressure of the incoming plasma. We
can thus assume that the overlapping region remains at roughly the
same size $L$ until $t=\tau_f$.

Since it took a time $\tau_s$ for the density of the central region of
size $L$ to increase from $n_0$ to $2n_0$, therefore, it should take
an additional time of $\tau_s$ to build the density up to the $3n_0$
required for the shock. Thus, we estimate the shock formation time $\tau_f$
to be given by
\begin{equation}\label{eq:form-time}
2D:\quad \tau_f=2\tau_s=\frac{1}{\delta}\ln \Pi.
\end{equation}

\begin{figure}[tb]
\begin{center}
\includegraphics[width=0.45\textwidth]{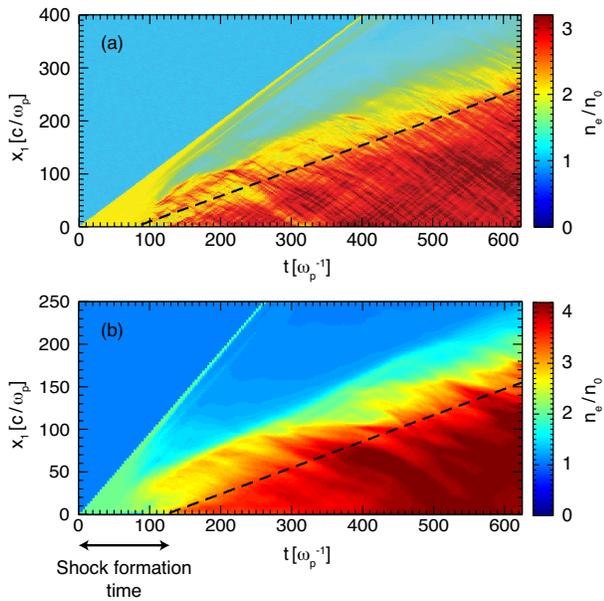}
\end{center}
\caption{(Color Online) Determination of the shock formation time for $\gamma_0=25$  for 2D (a) and 3D PIC simulations (b). Color indicates the density of the
  plasma (in units of $n_0$) as a function of time along the
  horizontal axis and position along the vertical axis. The dashed line delineates the boundary of the
  shocked plasma. The shock formation time is defined by the intersection
  of this line with the horizontal axis.} \label{fig:time}
\end{figure}

As a check of the above prediction, we measured the shock formation
time in a number of simulations using the method illustrated in
Fig.~\ref{fig:time}: we plot the transversely averaged density in the
simulation box as a function of time along the horizontal axis and
position along the vertical axis. The shock front appears clearly as
the boundary of the triangle of large densities
in the lower right corner of the plot. The slope of the hypotenuse,
shown by the dashed line, gives the speed of the shock front. The intersection of the
dashed line with the time axis corresponds to the shock formation time
$\tau_f \sim 85-90 \omega_p^{-1}$.  As an alternative shock formation criterion, we also
identified the moment when the density first reaches the expected
jump. Both methods yield very similar results.

In Fig.~\ref{fig:time}, there is a thin precursor escaping ahead of
the shock front with $v\sim c$. It corresponds to the dwindling
remnant of the border of the rightward-moving shell. Because this
narrow shell spans only a few $c/\omega_p$, it does not interact with
the upstream like a beam. Propagation is stable, preventing energy
dissipation through instabilities. Instead, the energy loss occurs
through the Bethe-like stopping process \cite{Bethe1930,jackson1998}, several orders of magnitudes slower than the Weibel
instability which is the primary cause of the shock
\cite{Niu1981}. The precursor cruises ahead of the front and becomes
quickly irrelevant to the shock formation and propagation.

In the case of a relativistic 3D system, the adiabatic index is
$\Gamma=4/3$, resulting in a density ratio in Eq. (\ref{eq:jump}) of $\sim
4$. Correspondingly, the shock formation time $\tau_f$ is estimated to
be
\begin{equation}\label{eq:form-time3}
3D:\quad \tau_f=3\tau_s=\frac{3}{2\delta}\ln \Pi.
\end{equation}

\begin{figure}[tb]
\begin{center}
\includegraphics[width=0.4\textwidth]{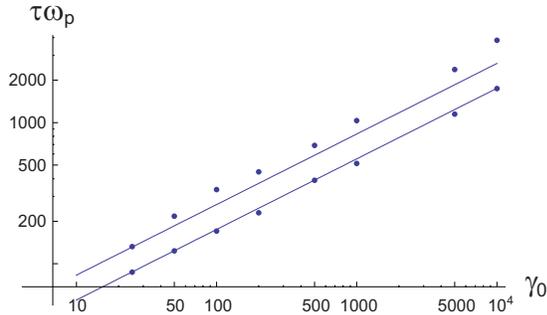}
\end{center}
\caption{Shock formation time predicted by Eq. (\ref{eq:form-time}) in 2D (lower line) and by Eq. (\ref{eq:form-time3}) in 3D (upper line), vs. simulation data (lower and upper dots, respectively).} \label{fig:form}
\end{figure}

Fig.~\ref{fig:form} compares the shock formation times estimated from
a number of 2D and 3D simulations spanning a wide range of $\gamma_0$ from 10
to $10^4$ with the theoretical predictions $\tau_f=2\tau_s$
(Eq. \ref{eq:form-time}) for the 2D case, and  $\tau_f=3\tau_s$
(Eq. \ref{eq:form-time3}) for the 3D case. Simulations and theory are in good agreement.

\section{Discussion}
From the present theory, the formation time of a Weibel-mediated shock can be derived. The analysis of the particle trajectories in the self-generated field shows that deviations from the forward directed motion occur only after field saturation. This will eventually lead to the energy dissipation and density accumulation in the shock front region. The time to achieve the steady-state shock compression ratio is twice the saturation time of the Weibel instability in 2D, $\tau_f = 2\tau_s$, and $\tau_f =3\tau_s$ in 3D. The comparison with 2D and 3D particle-in-cell simulations confirms this result.

We show that a shock always forms in a symmetric system of relativistically counterstreaming cold flows. Note however that our analysis holds provided the Weibel instability governs the unstable spectrum, which implies in turn $\gamma_0 > \sqrt{3/2}$, namely $\beta > 1/\sqrt{3}$ \cite{BretPoP2013}. It is expected that the shortest shock formation time is present at the maximum for the growth rate at $\gamma_0 = \sqrt{3}$ \cite{BretPoPReview}, which we plan to investigate in a fully relativistic approach in a following project.

The scenario exposed here is robust enough to be generalized to electron/ion plasma shocks which are currently under experimental scrutiny \cite{ross2012,Fox2013,Stockem2013,Huntington2013}. Here, electronic instabilities grow and saturate first on their proper time scale. The counter-streaming protons then propagate through the isotropized hot electrons and turn
unstable. Though the electron heating mechanism is yet to be fully understood \cite{Gedalin2012}, the filamentation instability of the proton beams should be dominant \cite{Yalinewich,Shaisultanov2012} and
reaches its maximum also for $\gamma_0 = \sqrt{3}$. Magnetic turbulence in the overlapping region will follow, resulting in the formation of a shock provided conditions similar to the present ones are fulfilled.

Noteworthily,  the choice of the reflecting wall technique forbids investigating collisions of plasma shells with different densities, compositions, or temperatures. Our present goal was to investigate the simplest possible system, in order to be able to put our formation scenario to the most accurate analytical test. Simulations of the full setup, without mirror, were performed where the same results have been obtained. The reflecting mirror has been used in order to save simulation time (factor 1/2) for the parameter study. Future works clearly contemplate studying asymmetric collisions, where the reflecting wall technique will be impossible to use.

\section{Acknowledgments}
This work was supported by the European Research Council (ERC-2010-AdG
Grant 267841) and FCT (Portugal) Grant Nos. PTDC/FIS/111720/2009 and
SFRH/BD/38952/2007. We acknowledge PRACE for providing access to resource SuperMUC based in Germany at the Leibniz research center. Thanks are due to Martin Lemoine, Laurent
Gremillet and Charles Ruyer for enriching discussions.


\begin{thebibliography}{43}%
\makeatletter
\providecommand \@ifxundefined [1]{%
 \@ifx{#1\undefined}
}%
\providecommand \@ifnum [1]{%
 \ifnum #1\expandafter \@firstoftwo
 \else \expandafter \@secondoftwo
 \fi
}%
\providecommand \@ifx [1]{%
 \ifx #1\expandafter \@firstoftwo
 \else \expandafter \@secondoftwo
 \fi
}%
\providecommand \natexlab [1]{#1}%
\providecommand \enquote  [1]{``#1''}%
\providecommand \bibnamefont  [1]{#1}%
\providecommand \bibfnamefont [1]{#1}%
\providecommand \citenamefont [1]{#1}%
\providecommand \href@noop [0]{\@secondoftwo}%
\providecommand \href [0]{\begingroup \@sanitize@url \@href}%
\providecommand \@href[1]{\@@startlink{#1}\@@href}%
\providecommand \@@href[1]{\endgroup#1\@@endlink}%
\providecommand \@sanitize@url [0]{\catcode `\\12\catcode `\$12\catcode
  `\&12\catcode `\#12\catcode `\^12\catcode `\_12\catcode `\%12\relax}%
\providecommand \@@startlink[1]{}%
\providecommand \@@endlink[0]{}%
\providecommand \url  [0]{\begingroup\@sanitize@url \@url }%
\providecommand \@url [1]{\endgroup\@href {#1}{\urlprefix }}%
\providecommand \urlprefix  [0]{URL }%
\providecommand \Eprint [0]{\href }%
\providecommand \doibase [0]{http://dx.doi.org/}%
\providecommand \selectlanguage [0]{\@gobble}%
\providecommand \bibinfo  [0]{\@secondoftwo}%
\providecommand \bibfield  [0]{\@secondoftwo}%
\providecommand \translation [1]{[#1]}%
\providecommand \BibitemOpen [0]{}%
\providecommand \bibitemStop [0]{}%
\providecommand \bibitemNoStop [0]{.\EOS\space}%
\providecommand \EOS [0]{\spacefactor3000\relax}%
\providecommand \BibitemShut  [1]{\csname bibitem#1\endcsname}%
\let\auto@bib@innerbib\@empty
\bibitem [{\citenamefont {Zel'dovich}\ and\ \citenamefont
  {Raizer}(2002)}]{Zeldovich}%
  \BibitemOpen
  \bibfield  {author} {\bibinfo {author} {\bibfnamefont {Y.~B.}\ \bibnamefont
  {Zel'dovich}}\ and\ \bibinfo {author} {\bibfnamefont {Y.~P.}\ \bibnamefont
  {Raizer}},\ }\href@noop {} {\emph {\bibinfo {title} {Physics of shock waves
  and high-temperature hydrodynamic phenomena}}},\ edited by\ \bibinfo {editor}
  {\bibfnamefont {W.~D.}\ \bibnamefont {Hayes}}\ and\ \bibinfo {editor}
  {\bibfnamefont {R.~F.}\ \bibnamefont {Probstein}}\ (\bibinfo  {publisher}
  {Dover Publications},\ \bibinfo {year} {2002})\ pp.\ \bibinfo {pages} {xxvii,
  916 p.}\BibitemShut {Stop}%
\bibitem [{\citenamefont {Bale}\ \emph {et~al.}(2003)\citenamefont {Bale},
  \citenamefont {Mozer},\ and\ \citenamefont {Horbury}}]{PRLBow1}%
  \BibitemOpen
  \bibfield  {author} {\bibinfo {author} {\bibfnamefont {S.~D.}\ \bibnamefont
  {Bale}}, \bibinfo {author} {\bibfnamefont {F.~S.}\ \bibnamefont {Mozer}}, \
  and\ \bibinfo {author} {\bibfnamefont {T.~S.}\ \bibnamefont {Horbury}},\
  }\href@noop {} {\bibfield  {journal} {\bibinfo  {journal} {Phys. Rev. Lett.}\
  }\textbf {\bibinfo {volume} {91}},\ \bibinfo {pages} {265004} (\bibinfo
  {year} {2003})}\BibitemShut {NoStop}%
\bibitem [{\citenamefont {Schwartz}\ \emph {et~al.}(2011)\citenamefont
  {Schwartz}, \citenamefont {Henley}, \citenamefont {Mitchell},\ and\
  \citenamefont {Krasnoselskikh}}]{PRLBow2}%
  \BibitemOpen
  \bibfield  {author} {\bibinfo {author} {\bibfnamefont {S.~J.}\ \bibnamefont
  {Schwartz}}, \bibinfo {author} {\bibfnamefont {E.}~\bibnamefont {Henley}},
  \bibinfo {author} {\bibfnamefont {J.}~\bibnamefont {Mitchell}}, \ and\
  \bibinfo {author} {\bibfnamefont {V.}~\bibnamefont {Krasnoselskikh}},\
  }\href@noop {} {\bibfield  {journal} {\bibinfo  {journal} {Phys. Rev. Lett.}\
  }\textbf {\bibinfo {volume} {107}},\ \bibinfo {pages} {215002} (\bibinfo
  {year} {2011})}\BibitemShut {NoStop}%
\bibitem [{\citenamefont {Kasper}\ \emph {et~al.}(2008)\citenamefont {Kasper},
  \citenamefont {Lazarus},\ and\ \citenamefont {Gary}}]{kasper2008}%
  \BibitemOpen
  \bibfield  {author} {\bibinfo {author} {\bibfnamefont {J.~C.}\ \bibnamefont
  {Kasper}}, \bibinfo {author} {\bibfnamefont {A.~J.}\ \bibnamefont {Lazarus}},
  \ and\ \bibinfo {author} {\bibfnamefont {S.~P.}\ \bibnamefont {Gary}},\
  }\href@noop {} {\bibfield  {journal} {\bibinfo  {journal} {Phys. Rev. Lett.}\
  }\textbf {\bibinfo {volume} {101}},\ \bibinfo {pages} {261103} (\bibinfo
  {year} {2008})}\BibitemShut {NoStop}%
\bibitem [{\citenamefont {Blandford}\ and\ \citenamefont
  {Ostriker}(1978)}]{Blandford78}%
  \BibitemOpen
  \bibfield  {author} {\bibinfo {author} {\bibfnamefont {R.}~\bibnamefont
  {Blandford}}\ and\ \bibinfo {author} {\bibfnamefont {J.}~\bibnamefont
  {Ostriker}},\ }\href@noop {} {\bibfield  {journal} {\bibinfo  {journal}
  {Astrophysical Journal}\ }\textbf {\bibinfo {volume} {221}},\ \bibinfo
  {pages} {L29} (\bibinfo {year} {1978})}\BibitemShut {NoStop}%
\bibitem [{\citenamefont {{Bell}}(1978{\natexlab{a}})}]{Bell1978a}%
  \BibitemOpen
  \bibfield  {author} {\bibinfo {author} {\bibfnamefont {A.~R.}\ \bibnamefont
  {{Bell}}},\ }\href@noop {} {\bibfield  {journal} {\bibinfo  {journal} {Mon.
  Not. R. Astron. Soc}\ }\textbf {\bibinfo {volume} {182}},\ \bibinfo {pages}
  {147} (\bibinfo {year} {1978}{\natexlab{a}})}\BibitemShut {NoStop}%
\bibitem [{\citenamefont {{Bell}}(1978{\natexlab{b}})}]{Bell1978b}%
  \BibitemOpen
  \bibfield  {author} {\bibinfo {author} {\bibfnamefont {A.~R.}\ \bibnamefont
  {{Bell}}},\ }\href@noop {} {\bibfield  {journal} {\bibinfo  {journal} {Mon.
  Not. R. Astron. Soc}\ }\textbf {\bibinfo {volume} {182}},\ \bibinfo {pages}
  {443} (\bibinfo {year} {1978}{\natexlab{b}})}\BibitemShut {NoStop}%
\bibitem [{\citenamefont {Spitkovsky}(2008)}]{Spitkovsky2008a}%
  \BibitemOpen
  \bibfield  {author} {\bibinfo {author} {\bibfnamefont {A.}~\bibnamefont
  {Spitkovsky}},\ }\href@noop {} {\bibfield  {journal} {\bibinfo  {journal}
  {Astrophys. J. Lett.}\ }\textbf {\bibinfo {volume} {682}},\ \bibinfo {pages}
  {L5} (\bibinfo {year} {2008})}\BibitemShut {NoStop}%
\bibitem [{\citenamefont {{Martins}}\ \emph {et~al.}(2009)\citenamefont
  {{Martins}}, \citenamefont {{Fonseca}}, \citenamefont {{Silva}},\ and\
  \citenamefont {{Mori}}}]{Martins2009}%
  \BibitemOpen
  \bibfield  {author} {\bibinfo {author} {\bibfnamefont {S.~F.}\ \bibnamefont
  {{Martins}}}, \bibinfo {author} {\bibfnamefont {R.~A.}\ \bibnamefont
  {{Fonseca}}}, \bibinfo {author} {\bibfnamefont {L.~O.}\ \bibnamefont
  {{Silva}}}, \ and\ \bibinfo {author} {\bibfnamefont {W.~B.}\ \bibnamefont
  {{Mori}}},\ }\href@noop {} {\bibfield  {journal} {\bibinfo  {journal}
  {Astrophysical Journal Letters}\ }\textbf {\bibinfo {volume} {695}},\
  \bibinfo {pages} {L189} (\bibinfo {year} {2009})}\BibitemShut {NoStop}%
\bibitem [{\citenamefont {Blandford}\ and\ \citenamefont
  {Eichler}(1987)}]{Blandford1987}%
  \BibitemOpen
  \bibfield  {author} {\bibinfo {author} {\bibfnamefont {R.}~\bibnamefont
  {Blandford}}\ and\ \bibinfo {author} {\bibfnamefont {D.}~\bibnamefont
  {Eichler}},\ }\href@noop {} {\bibfield  {journal} {\bibinfo  {journal} {Phys.
  Rep.}\ }\textbf {\bibinfo {volume} {154}},\ \bibinfo {pages} {1} (\bibinfo
  {year} {1987})}\BibitemShut {NoStop}%
\bibitem [{\citenamefont {Piran}(2004)}]{Piran2004}%
  \BibitemOpen
  \bibfield  {author} {\bibinfo {author} {\bibfnamefont {T.}~\bibnamefont
  {Piran}},\ }\href@noop {} {\bibfield  {journal} {\bibinfo  {journal} {Rev.
  Mod. Phys.}\ }\textbf {\bibinfo {volume} {76}},\ \bibinfo {pages} {1143}
  (\bibinfo {year} {2004})}\BibitemShut {NoStop}%
\bibitem [{\citenamefont {{Sagdeev}}(1966)}]{Sagdeev66}%
  \BibitemOpen
  \bibfield  {author} {\bibinfo {author} {\bibfnamefont {R.~Z.}\ \bibnamefont
  {{Sagdeev}}},\ }\href@noop {} {\bibfield  {journal} {\bibinfo  {journal}
  {Reviews of Plasma Physics}\ }\textbf {\bibinfo {volume} {4}},\ \bibinfo
  {pages} {23} (\bibinfo {year} {1966})}\BibitemShut {NoStop}%
\bibitem [{\citenamefont {{Spitkovsky}}(2005)}]{Spitkovsky2005}%
  \BibitemOpen
  \bibfield  {author} {\bibinfo {author} {\bibfnamefont {A.}~\bibnamefont
  {{Spitkovsky}}},\ }in\ \href {\doibase 10.1063/1.2141897} {\emph {\bibinfo
  {booktitle} {Astrophysical Sources of High Energy Particles and
  Radiation}}},\ \bibinfo {series} {American Institute of Physics Conference
  Series}, Vol.\ \bibinfo {volume} {801},\ \bibinfo {editor} {edited by\
  \bibinfo {editor} {\bibfnamefont {T.}~\bibnamefont {{Bulik}}}, \bibinfo
  {editor} {\bibfnamefont {B.}~\bibnamefont {{Rudak}}}, \ and\ \bibinfo
  {editor} {\bibfnamefont {G.}~\bibnamefont {{Madejski}}}}\ (\bibinfo {year}
  {2005})\ pp.\ \bibinfo {pages} {345--350},\ \Eprint
  {http://arxiv.org/abs/arXiv:astro-ph/0603211} {arXiv:astro-ph/0603211}
  \BibitemShut {NoStop}%
\bibitem [{\citenamefont {Kato}(2007)}]{kato2007}%
  \BibitemOpen
  \bibfield  {author} {\bibinfo {author} {\bibfnamefont {T.~N.}\ \bibnamefont
  {Kato}},\ }\href@noop {} {\bibfield  {journal} {\bibinfo  {journal} {The
  Astrophysical Journal}\ }\textbf {\bibinfo {volume} {668}},\ \bibinfo {pages}
  {974} (\bibinfo {year} {2007})}\BibitemShut {NoStop}%
\bibitem [{\citenamefont {Chang}\ \emph {et~al.}(2008)\citenamefont {Chang},
  \citenamefont {Spitkovsky},\ and\ \citenamefont {Arons}}]{Chang2008}%
  \BibitemOpen
  \bibfield  {author} {\bibinfo {author} {\bibfnamefont {P.}~\bibnamefont
  {Chang}}, \bibinfo {author} {\bibfnamefont {A.}~\bibnamefont {Spitkovsky}}, \
  and\ \bibinfo {author} {\bibfnamefont {J.}~\bibnamefont {Arons}},\
  }\href@noop {} {\bibfield  {journal} {\bibinfo  {journal} {The Astrophysical
  Journal}\ }\textbf {\bibinfo {volume} {674}},\ \bibinfo {pages} {378}
  (\bibinfo {year} {2008})}\BibitemShut {NoStop}%
\bibitem [{\citenamefont {Silva}\ \emph {et~al.}(2003)\citenamefont {Silva},
  \citenamefont {Fonseca}, \citenamefont {Tonge}, \citenamefont {Dawson},
  \citenamefont {Mori},\ and\ \citenamefont {Medvedev}}]{SilvaApJ}%
  \BibitemOpen
  \bibfield  {author} {\bibinfo {author} {\bibfnamefont {L.~O.}\ \bibnamefont
  {Silva}}, \bibinfo {author} {\bibfnamefont {R.~A.}\ \bibnamefont {Fonseca}},
  \bibinfo {author} {\bibfnamefont {J.~W.}\ \bibnamefont {Tonge}}, \bibinfo
  {author} {\bibfnamefont {J.~M.}\ \bibnamefont {Dawson}}, \bibinfo {author}
  {\bibfnamefont {W.~B.}\ \bibnamefont {Mori}}, \ and\ \bibinfo {author}
  {\bibfnamefont {M.~V.}\ \bibnamefont {Medvedev}},\ }\href@noop {} {\bibfield
  {journal} {\bibinfo  {journal} {Astrophys. J.}\ }\textbf {\bibinfo {volume}
  {596}},\ \bibinfo {pages} {L121} (\bibinfo {year} {2003})}\BibitemShut
  {NoStop}%
\bibitem [{\citenamefont {Bret}\ \emph {et~al.}(2013)\citenamefont {Bret},
  \citenamefont {Stockem}, \citenamefont {Fi\'{u}za}, \citenamefont {Ruyer},
  \citenamefont {Gremillet}, \citenamefont {Narayan},\ and\ \citenamefont
  {Silva}}]{BretPoP2013}%
  \BibitemOpen
  \bibfield  {author} {\bibinfo {author} {\bibfnamefont {A.}~\bibnamefont
  {Bret}}, \bibinfo {author} {\bibfnamefont {A.}~\bibnamefont {Stockem}},
  \bibinfo {author} {\bibfnamefont {F.}~\bibnamefont {Fi\'{u}za}}, \bibinfo
  {author} {\bibfnamefont {C.}~\bibnamefont {Ruyer}}, \bibinfo {author}
  {\bibfnamefont {L.}~\bibnamefont {Gremillet}}, \bibinfo {author}
  {\bibfnamefont {R.}~\bibnamefont {Narayan}}, \ and\ \bibinfo {author}
  {\bibfnamefont {L.~O.}\ \bibnamefont {Silva}},\ }\href@noop {} {\bibfield
  {journal} {\bibinfo  {journal} {Physics of Plasmas}\ }\textbf {\bibinfo
  {volume} {20}},\ \bibinfo {eid} {042102} (\bibinfo {year}
  {2013})}\BibitemShut {NoStop}%
\bibitem [{\citenamefont {Medvedev}\ and\ \citenamefont
  {Loeb}(1999)}]{Medvedev1999}%
  \BibitemOpen
  \bibfield  {author} {\bibinfo {author} {\bibfnamefont {M.~V.}\ \bibnamefont
  {Medvedev}}\ and\ \bibinfo {author} {\bibfnamefont {A.}~\bibnamefont
  {Loeb}},\ }\href@noop {} {\bibfield  {journal} {\bibinfo  {journal}
  {Astrophys. J.}\ }\textbf {\bibinfo {volume} {526}},\ \bibinfo {pages} {697}
  (\bibinfo {year} {1999})}\BibitemShut {NoStop}%
\bibitem [{\citenamefont {Ross}\ \emph {et~al.}(2012)\citenamefont {Ross},
  \citenamefont {Glenzer}, \citenamefont {Amendt}, \citenamefont {Berger},
  \citenamefont {Divol}, \citenamefont {Kugland}, \citenamefont {Landen},
  \citenamefont {Plechaty}, \citenamefont {Remington}, \citenamefont {Ryutov},
  \citenamefont {Rozmus}, \citenamefont {Froula}, \citenamefont {Fiksel},
  \citenamefont {Sorce}, \citenamefont {Kuramitsu}, \citenamefont {Morita},
  \citenamefont {Sakawa}, \citenamefont {Takabe}, \citenamefont {Drake},
  \citenamefont {Grosskopf}, \citenamefont {Kuranz}, \citenamefont {Gregori},
  \citenamefont {Meinecke}, \citenamefont {Murphy}, \citenamefont {Koenig},
  \citenamefont {Pelka}, \citenamefont {Ravasio}, \citenamefont {Vinci},
  \citenamefont {Liang}, \citenamefont {Presura}, \citenamefont {Spitkovsky},
  \citenamefont {Miniati},\ and\ \citenamefont {Park}}]{ross2012}%
  \BibitemOpen
  \bibfield  {author} {\bibinfo {author} {\bibfnamefont {J.~S.}\ \bibnamefont
  {Ross}}, \bibinfo {author} {\bibfnamefont {S.~H.}\ \bibnamefont {Glenzer}},
  \bibinfo {author} {\bibfnamefont {P.}~\bibnamefont {Amendt}}, \bibinfo
  {author} {\bibfnamefont {R.}~\bibnamefont {Berger}}, \bibinfo {author}
  {\bibfnamefont {L.}~\bibnamefont {Divol}}, \bibinfo {author} {\bibfnamefont
  {N.~L.}\ \bibnamefont {Kugland}}, \bibinfo {author} {\bibfnamefont {O.~L.}\
  \bibnamefont {Landen}}, \bibinfo {author} {\bibfnamefont {C.}~\bibnamefont
  {Plechaty}}, \bibinfo {author} {\bibfnamefont {B.}~\bibnamefont {Remington}},
  \bibinfo {author} {\bibfnamefont {D.}~\bibnamefont {Ryutov}}, \bibinfo
  {author} {\bibfnamefont {W.}~\bibnamefont {Rozmus}}, \bibinfo {author}
  {\bibfnamefont {D.~H.}\ \bibnamefont {Froula}}, \bibinfo {author}
  {\bibfnamefont {G.}~\bibnamefont {Fiksel}}, \bibinfo {author} {\bibfnamefont
  {C.}~\bibnamefont {Sorce}}, \bibinfo {author} {\bibfnamefont
  {Y.}~\bibnamefont {Kuramitsu}}, \bibinfo {author} {\bibfnamefont
  {T.}~\bibnamefont {Morita}}, \bibinfo {author} {\bibfnamefont
  {Y.}~\bibnamefont {Sakawa}}, \bibinfo {author} {\bibfnamefont
  {H.}~\bibnamefont {Takabe}}, \bibinfo {author} {\bibfnamefont {R.~P.}\
  \bibnamefont {Drake}}, \bibinfo {author} {\bibfnamefont {M.}~\bibnamefont
  {Grosskopf}}, \bibinfo {author} {\bibfnamefont {C.}~\bibnamefont {Kuranz}},
  \bibinfo {author} {\bibfnamefont {G.}~\bibnamefont {Gregori}}, \bibinfo
  {author} {\bibfnamefont {J.}~\bibnamefont {Meinecke}}, \bibinfo {author}
  {\bibfnamefont {C.~D.}\ \bibnamefont {Murphy}}, \bibinfo {author}
  {\bibfnamefont {M.}~\bibnamefont {Koenig}}, \bibinfo {author} {\bibfnamefont
  {A.}~\bibnamefont {Pelka}}, \bibinfo {author} {\bibfnamefont
  {A.}~\bibnamefont {Ravasio}}, \bibinfo {author} {\bibfnamefont
  {T.}~\bibnamefont {Vinci}}, \bibinfo {author} {\bibfnamefont
  {E.}~\bibnamefont {Liang}}, \bibinfo {author} {\bibfnamefont
  {R.}~\bibnamefont {Presura}}, \bibinfo {author} {\bibfnamefont
  {A.}~\bibnamefont {Spitkovsky}}, \bibinfo {author} {\bibfnamefont
  {F.}~\bibnamefont {Miniati}}, \ and\ \bibinfo {author} {\bibfnamefont
  {H.-S.}\ \bibnamefont {Park}},\ }\href@noop {} {\bibfield  {journal}
  {\bibinfo  {journal} {Physics of Plasmas}\ }\textbf {\bibinfo {volume}
  {19}},\ \bibinfo {pages} {056501} (\bibinfo {year} {2012})}\BibitemShut
  {NoStop}%
\bibitem [{\citenamefont {{Kugland}}\ \emph {et~al.}(2012)\citenamefont
  {{Kugland}}, \citenamefont {{Ryutov}}, \citenamefont {{Chang}}, \citenamefont
  {{Drake}}, \citenamefont {{Fiksel}}, \citenamefont {{Froula}}, \citenamefont
  {{Glenzer}}, \citenamefont {{Gregori}}, \citenamefont {{Grosskopf}},
  \citenamefont {{Koenig}}, \citenamefont {{Kuramitsu}}, \citenamefont
  {{Kuranz}}, \citenamefont {{Levy}}, \citenamefont {{Liang}}, \citenamefont
  {{Meinecke}}, \citenamefont {{Miniati}}, \citenamefont {{Morita}},
  \citenamefont {{Pelka}}, \citenamefont {{Plechaty}}, \citenamefont
  {{Presura}}, \citenamefont {{Ravasio}}, \citenamefont {{Remington}},
  \citenamefont {{Reville}}, \citenamefont {{Ross}}, \citenamefont {{Sakawa}},
  \citenamefont {{Spitkovsky}}, \citenamefont {{Takabe}},\ and\ \citenamefont
  {{Park}}}]{Kugland2012}%
  \BibitemOpen
  \bibfield  {author} {\bibinfo {author} {\bibfnamefont {N.~L.}\ \bibnamefont
  {{Kugland}}}, \bibinfo {author} {\bibfnamefont {D.~D.}\ \bibnamefont
  {{Ryutov}}}, \bibinfo {author} {\bibfnamefont {P.-Y.}\ \bibnamefont
  {{Chang}}}, \bibinfo {author} {\bibfnamefont {R.~P.}\ \bibnamefont
  {{Drake}}}, \bibinfo {author} {\bibfnamefont {G.}~\bibnamefont {{Fiksel}}},
  \bibinfo {author} {\bibfnamefont {D.~H.}\ \bibnamefont {{Froula}}}, \bibinfo
  {author} {\bibfnamefont {S.~H.}\ \bibnamefont {{Glenzer}}}, \bibinfo {author}
  {\bibfnamefont {G.}~\bibnamefont {{Gregori}}}, \bibinfo {author}
  {\bibfnamefont {M.}~\bibnamefont {{Grosskopf}}}, \bibinfo {author}
  {\bibfnamefont {M.}~\bibnamefont {{Koenig}}}, \bibinfo {author}
  {\bibfnamefont {Y.}~\bibnamefont {{Kuramitsu}}}, \bibinfo {author}
  {\bibfnamefont {C.}~\bibnamefont {{Kuranz}}}, \bibinfo {author}
  {\bibfnamefont {M.~C.}\ \bibnamefont {{Levy}}}, \bibinfo {author}
  {\bibfnamefont {E.}~\bibnamefont {{Liang}}}, \bibinfo {author} {\bibfnamefont
  {J.}~\bibnamefont {{Meinecke}}}, \bibinfo {author} {\bibfnamefont
  {F.}~\bibnamefont {{Miniati}}}, \bibinfo {author} {\bibfnamefont
  {T.}~\bibnamefont {{Morita}}}, \bibinfo {author} {\bibfnamefont
  {A.}~\bibnamefont {{Pelka}}}, \bibinfo {author} {\bibfnamefont
  {C.}~\bibnamefont {{Plechaty}}}, \bibinfo {author} {\bibfnamefont
  {R.}~\bibnamefont {{Presura}}}, \bibinfo {author} {\bibfnamefont
  {A.}~\bibnamefont {{Ravasio}}}, \bibinfo {author} {\bibfnamefont {B.~A.}\
  \bibnamefont {{Remington}}}, \bibinfo {author} {\bibfnamefont
  {B.}~\bibnamefont {{Reville}}}, \bibinfo {author} {\bibfnamefont {J.~S.}\
  \bibnamefont {{Ross}}}, \bibinfo {author} {\bibfnamefont {Y.}~\bibnamefont
  {{Sakawa}}}, \bibinfo {author} {\bibfnamefont {A.}~\bibnamefont
  {{Spitkovsky}}}, \bibinfo {author} {\bibfnamefont {H.}~\bibnamefont
  {{Takabe}}}, \ and\ \bibinfo {author} {\bibfnamefont {H.-S.}\ \bibnamefont
  {{Park}}},\ }\href {\doibase 10.1038/nphys2434} {\bibfield  {journal}
  {\bibinfo  {journal} {Nature Physics}\ }\textbf {\bibinfo {volume} {8}},\
  \bibinfo {pages} {809} (\bibinfo {year} {2012})}\BibitemShut {NoStop}%
\bibitem [{\citenamefont {Fox}\ \emph {et~al.}(2013)\citenamefont {Fox},
  \citenamefont {Fiksel}, \citenamefont {Bhattacharjee}, \citenamefont {Chang},
  \citenamefont {Germaschewski}, \citenamefont {Hu},\ and\ \citenamefont
  {Nilson}}]{Fox2013}%
  \BibitemOpen
  \bibfield  {author} {\bibinfo {author} {\bibfnamefont {W.}~\bibnamefont
  {Fox}}, \bibinfo {author} {\bibfnamefont {G.}~\bibnamefont {Fiksel}},
  \bibinfo {author} {\bibfnamefont {A.}~\bibnamefont {Bhattacharjee}}, \bibinfo
  {author} {\bibfnamefont {P.-Y.}\ \bibnamefont {Chang}}, \bibinfo {author}
  {\bibfnamefont {K.}~\bibnamefont {Germaschewski}}, \bibinfo {author}
  {\bibfnamefont {S.~X.}\ \bibnamefont {Hu}}, \ and\ \bibinfo {author}
  {\bibfnamefont {P.~M.}\ \bibnamefont {Nilson}},\ }\href {\doibase
  10.1103/PhysRevLett.111.225002} {\bibfield  {journal} {\bibinfo  {journal}
  {Phys. Rev. Lett.}\ }\textbf {\bibinfo {volume} {111}},\ \bibinfo {pages}
  {225002} (\bibinfo {year} {2013})}\BibitemShut {NoStop}%
\bibitem [{\citenamefont {{Huntington}}\ \emph {et~al.}(2013)\citenamefont
  {{Huntington}}, \citenamefont {{Fi\'{u}za}}, \citenamefont {{Ross}},
  \citenamefont {{Zylstra}}, \citenamefont {{Drake}}, \citenamefont {{Froula}},
  \citenamefont {{Gregori}}, \citenamefont {{Kugland}}, \citenamefont
  {{Kuranz}}, \citenamefont {{Levy}}, \citenamefont {{Li}}, \citenamefont
  {{Meinecke}}, \citenamefont {{Morita}}, \citenamefont {{Petrasso}},
  \citenamefont {{Plechaty}}, \citenamefont {{Remington}}, \citenamefont
  {{Ryutov}}, \citenamefont {{Sakawa}}, \citenamefont {{Spitkovsky}},
  \citenamefont {{Takabe}},\ and\ \citenamefont {{Park}}}]{Huntington2013}%
  \BibitemOpen
  \bibfield  {author} {\bibinfo {author} {\bibfnamefont {C.~M.}\ \bibnamefont
  {{Huntington}}}, \bibinfo {author} {\bibfnamefont {F.}~\bibnamefont
  {{Fi\'{u}za}}}, \bibinfo {author} {\bibfnamefont {J.~S.}\ \bibnamefont
  {{Ross}}}, \bibinfo {author} {\bibfnamefont {A.~B.}\ \bibnamefont
  {{Zylstra}}}, \bibinfo {author} {\bibfnamefont {R.~P.}\ \bibnamefont
  {{Drake}}}, \bibinfo {author} {\bibfnamefont {D.~H.}\ \bibnamefont
  {{Froula}}}, \bibinfo {author} {\bibfnamefont {G.}~\bibnamefont {{Gregori}}},
  \bibinfo {author} {\bibfnamefont {N.~L.}\ \bibnamefont {{Kugland}}}, \bibinfo
  {author} {\bibfnamefont {C.~C.}\ \bibnamefont {{Kuranz}}}, \bibinfo {author}
  {\bibfnamefont {M.~C.}\ \bibnamefont {{Levy}}}, \bibinfo {author}
  {\bibfnamefont {C.~K.}\ \bibnamefont {{Li}}}, \bibinfo {author}
  {\bibfnamefont {J.}~\bibnamefont {{Meinecke}}}, \bibinfo {author}
  {\bibfnamefont {T.}~\bibnamefont {{Morita}}}, \bibinfo {author}
  {\bibfnamefont {R.}~\bibnamefont {{Petrasso}}}, \bibinfo {author}
  {\bibfnamefont {C.}~\bibnamefont {{Plechaty}}}, \bibinfo {author}
  {\bibfnamefont {B.~A.}\ \bibnamefont {{Remington}}}, \bibinfo {author}
  {\bibfnamefont {D.~D.}\ \bibnamefont {{Ryutov}}}, \bibinfo {author}
  {\bibfnamefont {Y.}~\bibnamefont {{Sakawa}}}, \bibinfo {author}
  {\bibfnamefont {A.}~\bibnamefont {{Spitkovsky}}}, \bibinfo {author}
  {\bibfnamefont {H.}~\bibnamefont {{Takabe}}}, \ and\ \bibinfo {author}
  {\bibfnamefont {H.-S.}\ \bibnamefont {{Park}}},\ }\href@noop {} {\  (\bibinfo
  {year} {2013})},\ \Eprint {http://arxiv.org/abs/arXiv:1310.3337}
  {arXiv:1310.3337} \BibitemShut {NoStop}%
\bibitem [{\citenamefont {{Muggli}}\ \emph {et~al.}(2013)\citenamefont
  {{Muggli}}, \citenamefont {{Martins}}, \citenamefont {{Vieira}},\ and\
  \citenamefont {{Silva}}}]{Muggli2013}%
  \BibitemOpen
  \bibfield  {author} {\bibinfo {author} {\bibfnamefont {P.}~\bibnamefont
  {{Muggli}}}, \bibinfo {author} {\bibfnamefont {S.~F.}\ \bibnamefont
  {{Martins}}}, \bibinfo {author} {\bibfnamefont {J.}~\bibnamefont {{Vieira}}},
  \ and\ \bibinfo {author} {\bibfnamefont {L.~O.}\ \bibnamefont {{Silva}}},\
  }\href@noop {} {\bibfield  {journal} {\bibinfo  {journal} {ArXiv e-prints}\ }
  (\bibinfo {year} {2013})},\ \Eprint {http://arxiv.org/abs/1306.4380}
  {arXiv:1306.4380} \BibitemShut {NoStop}%
\bibitem [{\citenamefont {Sarri}\ \emph {et~al.}(2013)\citenamefont {Sarri},
  \citenamefont {Schumaker}, \citenamefont {Di~Piazza}, \citenamefont {Vargas},
  \citenamefont {Dromey}, \citenamefont {Dieckmann}, \citenamefont {Chvykov},
  \citenamefont {Maksimchuk}, \citenamefont {Yanovsky}, \citenamefont {He},
  \citenamefont {Hou}, \citenamefont {Nees}, \citenamefont {Thomas},
  \citenamefont {Keitel}, \citenamefont {Zepf},\ and\ \citenamefont
  {Krushelnick}}]{Sarri2013}%
  \BibitemOpen
  \bibfield  {author} {\bibinfo {author} {\bibfnamefont {G.}~\bibnamefont
  {Sarri}}, \bibinfo {author} {\bibfnamefont {W.}~\bibnamefont {Schumaker}},
  \bibinfo {author} {\bibfnamefont {A.}~\bibnamefont {Di~Piazza}}, \bibinfo
  {author} {\bibfnamefont {M.}~\bibnamefont {Vargas}}, \bibinfo {author}
  {\bibfnamefont {B.}~\bibnamefont {Dromey}}, \bibinfo {author} {\bibfnamefont
  {M.~E.}\ \bibnamefont {Dieckmann}}, \bibinfo {author} {\bibfnamefont
  {V.}~\bibnamefont {Chvykov}}, \bibinfo {author} {\bibfnamefont
  {A.}~\bibnamefont {Maksimchuk}}, \bibinfo {author} {\bibfnamefont
  {V.}~\bibnamefont {Yanovsky}}, \bibinfo {author} {\bibfnamefont {Z.~H.}\
  \bibnamefont {He}}, \bibinfo {author} {\bibfnamefont {B.~X.}\ \bibnamefont
  {Hou}}, \bibinfo {author} {\bibfnamefont {J.~A.}\ \bibnamefont {Nees}},
  \bibinfo {author} {\bibfnamefont {A.~G.~R.}\ \bibnamefont {Thomas}}, \bibinfo
  {author} {\bibfnamefont {C.~H.}\ \bibnamefont {Keitel}}, \bibinfo {author}
  {\bibfnamefont {M.}~\bibnamefont {Zepf}}, \ and\ \bibinfo {author}
  {\bibfnamefont {K.}~\bibnamefont {Krushelnick}},\ }\href {\doibase
  10.1103/PhysRevLett.110.255002} {\bibfield  {journal} {\bibinfo  {journal}
  {Phys. Rev. Lett.}\ }\textbf {\bibinfo {volume} {110}},\ \bibinfo {pages}
  {255002} (\bibinfo {year} {2013})}\BibitemShut {NoStop}%
\bibitem [{\citenamefont {Stockem}\ \emph {et~al.}(2014)\citenamefont
  {Stockem}, \citenamefont {Fiuza}, \citenamefont {Bret}, \citenamefont
  {Fonseca},\ and\ \citenamefont {Silva}}]{Stockem2013}%
  \BibitemOpen
  \bibfield  {author} {\bibinfo {author} {\bibfnamefont {A.}~\bibnamefont
  {Stockem}}, \bibinfo {author} {\bibfnamefont {F.}~\bibnamefont {Fiuza}},
  \bibinfo {author} {\bibfnamefont {A.}~\bibnamefont {Bret}}, \bibinfo {author}
  {\bibfnamefont {R.}~\bibnamefont {Fonseca}}, \ and\ \bibinfo {author}
  {\bibfnamefont {L.}~\bibnamefont {Silva}},\ }\href@noop {} {\bibfield
  {journal} {\bibinfo  {journal} {Scientific Reports}\ }\textbf {\bibinfo
  {volume} {4}},\ \bibinfo {pages} {3934} (\bibinfo {year} {2014})}\BibitemShut
  {NoStop}%
\bibitem [{\citenamefont {Weibel}(1959)}]{Weibel}%
  \BibitemOpen
  \bibfield  {author} {\bibinfo {author} {\bibfnamefont {E.~S.}\ \bibnamefont
  {Weibel}},\ }\href@noop {} {\bibfield  {journal} {\bibinfo  {journal} {Phys.
  Rev. Lett.}\ }\textbf {\bibinfo {volume} {2}},\ \bibinfo {pages} {83}
  (\bibinfo {year} {1959})}\BibitemShut {NoStop}%
\bibitem [{\citenamefont {Fried}(1959)}]{Fried1959}%
  \BibitemOpen
  \bibfield  {author} {\bibinfo {author} {\bibfnamefont {B.~D.}\ \bibnamefont
  {Fried}},\ }\href@noop {} {\bibfield  {journal} {\bibinfo  {journal} {Phys.
  Fluids}\ }\textbf {\bibinfo {volume} {2}},\ \bibinfo {pages} {337} (\bibinfo
  {year} {1959})}\BibitemShut {NoStop}%
\bibitem [{\citenamefont {Deutsch}\ \emph {et~al.}(2005)\citenamefont
  {Deutsch}, \citenamefont {Bret}, \citenamefont {Firpo},\ and\ \citenamefont
  {Fromy}}]{DeutschPRE2005}%
  \BibitemOpen
  \bibfield  {author} {\bibinfo {author} {\bibfnamefont {C.}~\bibnamefont
  {Deutsch}}, \bibinfo {author} {\bibfnamefont {A.}~\bibnamefont {Bret}},
  \bibinfo {author} {\bibfnamefont {M.-C.}\ \bibnamefont {Firpo}}, \ and\
  \bibinfo {author} {\bibfnamefont {P.}~\bibnamefont {Fromy}},\ }\href@noop {}
  {\bibfield  {journal} {\bibinfo  {journal} {Phys. Rev. E}\ }\textbf {\bibinfo
  {volume} {72}},\ \bibinfo {pages} {026402} (\bibinfo {year}
  {2005})}\BibitemShut {NoStop}%
\bibitem [{\citenamefont {Bret}\ \emph {et~al.}(2010)\citenamefont {Bret},
  \citenamefont {Gremillet},\ and\ \citenamefont {Dieckmann}}]{BretPoPReview}%
  \BibitemOpen
  \bibfield  {author} {\bibinfo {author} {\bibfnamefont {A.}~\bibnamefont
  {Bret}}, \bibinfo {author} {\bibfnamefont {L.}~\bibnamefont {Gremillet}}, \
  and\ \bibinfo {author} {\bibfnamefont {M.~E.}\ \bibnamefont {Dieckmann}},\
  }\href@noop {} {\bibfield  {journal} {\bibinfo  {journal} {Phys. Plasmas}\
  }\textbf {\bibinfo {volume} {17}},\ \bibinfo {pages} {120501} (\bibinfo
  {year} {2010})}\BibitemShut {NoStop}%
\bibitem [{\citenamefont {Bret}\ \emph {et~al.}(2008)\citenamefont {Bret},
  \citenamefont {Gremillet}, \citenamefont {B\'{e}nisti},\ and\ \citenamefont
  {Lefebvre}}]{BretPRL2008}%
  \BibitemOpen
  \bibfield  {author} {\bibinfo {author} {\bibfnamefont {A.}~\bibnamefont
  {Bret}}, \bibinfo {author} {\bibfnamefont {L.}~\bibnamefont {Gremillet}},
  \bibinfo {author} {\bibfnamefont {D.}~\bibnamefont {B\'{e}nisti}}, \ and\
  \bibinfo {author} {\bibfnamefont {E.}~\bibnamefont {Lefebvre}},\ }\href@noop
  {} {\bibfield  {journal} {\bibinfo  {journal} {Phys. Rev. Lett.}\ }\textbf
  {\bibinfo {volume} {100}},\ \bibinfo {pages} {205008} (\bibinfo {year}
  {2008})}\BibitemShut {NoStop}%
\bibitem [{\citenamefont {Benford}(1973)}]{Benford1973}%
  \BibitemOpen
  \bibfield  {author} {\bibinfo {author} {\bibfnamefont {G.}~\bibnamefont
  {Benford}},\ }\href@noop {} {\bibfield  {journal} {\bibinfo  {journal}
  {Plasma Physics}\ }\textbf {\bibinfo {volume} {15}},\ \bibinfo {pages} {483}
  (\bibinfo {year} {1973})}\BibitemShut {NoStop}%
\bibitem [{\citenamefont {Califano}\ \emph {et~al.}(1998)\citenamefont
  {Califano}, \citenamefont {Prandi}, \citenamefont {Pegoraro},\ and\
  \citenamefont {Bulanov}}]{califano3}%
  \BibitemOpen
  \bibfield  {author} {\bibinfo {author} {\bibfnamefont {F.}~\bibnamefont
  {Califano}}, \bibinfo {author} {\bibfnamefont {R.}~\bibnamefont {Prandi}},
  \bibinfo {author} {\bibfnamefont {F.}~\bibnamefont {Pegoraro}}, \ and\
  \bibinfo {author} {\bibfnamefont {S.~V.}\ \bibnamefont {Bulanov}},\
  }\href@noop {} {\bibfield  {journal} {\bibinfo  {journal} {Phys. Rev. E}\
  }\textbf {\bibinfo {volume} {58}},\ \bibinfo {pages} {7837} (\bibinfo {year}
  {1998})}\BibitemShut {NoStop}%
\bibitem [{\citenamefont {Davidson}\ \emph {et~al.}(1972)\citenamefont
  {Davidson}, \citenamefont {Hammer}, \citenamefont {Haber},\ and\
  \citenamefont {Wagner}}]{davidsonPIC1972}%
  \BibitemOpen
  \bibfield  {author} {\bibinfo {author} {\bibfnamefont {R.~C.}\ \bibnamefont
  {Davidson}}, \bibinfo {author} {\bibfnamefont {D.~A.}\ \bibnamefont
  {Hammer}}, \bibinfo {author} {\bibfnamefont {I.}~\bibnamefont {Haber}}, \
  and\ \bibinfo {author} {\bibfnamefont {C.~E.}\ \bibnamefont {Wagner}},\
  }\href@noop {} {\bibfield  {journal} {\bibinfo  {journal} {Phys. Fluids}\
  }\textbf {\bibinfo {volume} {15}},\ \bibinfo {pages} {317} (\bibinfo {year}
  {1972})}\BibitemShut {NoStop}%
\bibitem [{\citenamefont {Sitenko}(1967)}]{Sitenko}%
  \BibitemOpen
  \bibfield  {author} {\bibinfo {author} {\bibfnamefont {A.~G.}\ \bibnamefont
  {Sitenko}},\ }\href@noop {} {\emph {\bibinfo {title} {Electromagnetic
  Fluctuations in Plasma}}}\ (\bibinfo  {publisher} {Academic Press, New
  York},\ \bibinfo {year} {1967})\BibitemShut {NoStop}%
\bibitem [{\citenamefont {{Ruyer}}\ \emph {et~al.}(2013)\citenamefont
  {{Ruyer}}, \citenamefont {{Gremillet}}, \citenamefont {{B{\'e}nisti}},\ and\
  \citenamefont {{Bonnaud}}}]{Ruyer2013}%
  \BibitemOpen
  \bibfield  {author} {\bibinfo {author} {\bibfnamefont {C.}~\bibnamefont
  {{Ruyer}}}, \bibinfo {author} {\bibfnamefont {L.}~\bibnamefont
  {{Gremillet}}}, \bibinfo {author} {\bibfnamefont {D.}~\bibnamefont
  {{B{\'e}nisti}}}, \ and\ \bibinfo {author} {\bibfnamefont {G.}~\bibnamefont
  {{Bonnaud}}},\ }\href@noop {} {\bibfield  {journal} {\bibinfo  {journal}
  {Phys. Plasmas}\ }\textbf {\bibinfo {volume} {20}},\ \bibinfo {pages}
  {112104} (\bibinfo {year} {2013})}\BibitemShut {NoStop}%
\bibitem [{\citenamefont {Blandford}\ and\ \citenamefont
  {McKee}(1976)}]{blandford76}%
  \BibitemOpen
  \bibfield  {author} {\bibinfo {author} {\bibfnamefont {R.~D.}\ \bibnamefont
  {Blandford}}\ and\ \bibinfo {author} {\bibfnamefont {C.~F.}\ \bibnamefont
  {McKee}},\ }\href@noop {} {\bibfield  {journal} {\bibinfo  {journal} {Physics
  of Fluids}\ }\textbf {\bibinfo {volume} {19}},\ \bibinfo {pages} {1130}
  (\bibinfo {year} {1976})}\BibitemShut {NoStop}%
\bibitem [{\citenamefont {Stockem}\ \emph {et~al.}(2012)\citenamefont
  {Stockem}, \citenamefont {Fi\'{u}za}, \citenamefont {Fonseca},\ and\
  \citenamefont {Silva}}]{Stockem2012}%
  \BibitemOpen
  \bibfield  {author} {\bibinfo {author} {\bibfnamefont {A.}~\bibnamefont
  {Stockem}}, \bibinfo {author} {\bibfnamefont {F.}~\bibnamefont {Fi\'{u}za}},
  \bibinfo {author} {\bibfnamefont {R.~A.}\ \bibnamefont {Fonseca}}, \ and\
  \bibinfo {author} {\bibfnamefont {L.~O.}\ \bibnamefont {Silva}},\ }\href@noop
  {} {\bibfield  {journal} {\bibinfo  {journal} {Plasma Physics and Controlled
  Fusion}\ }\textbf {\bibinfo {volume} {54}},\ \bibinfo {pages} {125004}
  (\bibinfo {year} {2012})}\BibitemShut {NoStop}%
\bibitem [{\citenamefont {Bethe}(1930)}]{Bethe1930}%
  \BibitemOpen
  \bibfield  {author} {\bibinfo {author} {\bibfnamefont {H.}~\bibnamefont
  {Bethe}},\ }\href@noop {} {\bibfield  {journal} {\bibinfo  {journal} {Annalen
  der Physik}\ }\textbf {\bibinfo {volume} {5}},\ \bibinfo {pages} {325}
  (\bibinfo {year} {1930})}\BibitemShut {NoStop}%
\bibitem [{\citenamefont {Jackson}(1998)}]{jackson1998}%
  \BibitemOpen
  \bibfield  {author} {\bibinfo {author} {\bibfnamefont {J.}~\bibnamefont
  {Jackson}},\ }\href@noop {} {\emph {\bibinfo {title} {Classical
  Electrodynamics}}}\ (\bibinfo  {publisher} {Wiley},\ \bibinfo {year}
  {1998})\BibitemShut {NoStop}%
\bibitem [{\citenamefont {Abe}\ and\ \citenamefont {Niu}(1981)}]{Niu1981}%
  \BibitemOpen
  \bibfield  {author} {\bibinfo {author} {\bibfnamefont {T.}~\bibnamefont
  {Abe}}\ and\ \bibinfo {author} {\bibfnamefont {K.}~\bibnamefont {Niu}},\
  }\href@noop {} {\bibfield  {journal} {\bibinfo  {journal} {Journal of the
  Physical Society of Japan}\ }\textbf {\bibinfo {volume} {50}},\ \bibinfo
  {pages} {949} (\bibinfo {year} {1981})}\BibitemShut {NoStop}%
\bibitem [{\citenamefont {{Gedalin}}\ \emph {et~al.}(2012)\citenamefont
  {{Gedalin}}, \citenamefont {{Smolik}}, \citenamefont {{Spitkovsky}},\ and\
  \citenamefont {{Balikhin}}}]{Gedalin2012}%
  \BibitemOpen
  \bibfield  {author} {\bibinfo {author} {\bibfnamefont {M.}~\bibnamefont
  {{Gedalin}}}, \bibinfo {author} {\bibfnamefont {E.}~\bibnamefont {{Smolik}}},
  \bibinfo {author} {\bibfnamefont {A.}~\bibnamefont {{Spitkovsky}}}, \ and\
  \bibinfo {author} {\bibfnamefont {M.}~\bibnamefont {{Balikhin}}},\ }\href
  {\doibase 10.1209/0295-5075/97/35002} {\bibfield  {journal} {\bibinfo
  {journal} {EPL (Europhysics Letters)}\ }\textbf {\bibinfo {volume} {97}},\
  \bibinfo {pages} {35002} (\bibinfo {year} {2012})}\BibitemShut {NoStop}%
\bibitem [{\citenamefont {Yalinewich}\ and\ \citenamefont
  {Gedalin}(2010)}]{Yalinewich}%
  \BibitemOpen
  \bibfield  {author} {\bibinfo {author} {\bibfnamefont {A.}~\bibnamefont
  {Yalinewich}}\ and\ \bibinfo {author} {\bibfnamefont {M.}~\bibnamefont
  {Gedalin}},\ }\href {\doibase http://dx.doi.org/10.1063/1.3432722} {\bibfield
   {journal} {\bibinfo  {journal} {Phys. Plasmas}\ }\textbf {\bibinfo {volume}
  {17}},\ \bibinfo {pages} {062101} (\bibinfo {year} {2010})}\BibitemShut
  {NoStop}%
\bibitem [{\citenamefont {Shaisultanov}\ \emph {et~al.}(2012)\citenamefont
  {Shaisultanov}, \citenamefont {Lyubarsky},\ and\ \citenamefont
  {Eichler}}]{Shaisultanov2012}%
  \BibitemOpen
  \bibfield  {author} {\bibinfo {author} {\bibfnamefont {R.}~\bibnamefont
  {Shaisultanov}}, \bibinfo {author} {\bibfnamefont {Y.}~\bibnamefont
  {Lyubarsky}}, \ and\ \bibinfo {author} {\bibfnamefont {D.}~\bibnamefont
  {Eichler}},\ }\href@noop {} {\bibfield  {journal} {\bibinfo  {journal} {The
  Astrophysical Journal}\ }\textbf {\bibinfo {volume} {744}},\ \bibinfo {pages}
  {182} (\bibinfo {year} {2012})}\BibitemShut {NoStop}%
\end{thebibliography}

%

\end{document}